\documentclass[12pt,a4paper]{article}

\usepackage{amsmath}
\usepackage{graphicx}
\usepackage{times}
\usepackage{cite}
\usepackage{ulem}
\begin{document}
\begin{titlepage}
\title{Multiplicity distribution under the reflective scattering mode}
\author{ S.M. Troshin, N.E. Tyurin\\[1ex]
\small  \it NRC ``Kurchatov Institute''--IHEP\\
\small  \it Protvino, 142281, Russian Federation,\\
\small Sergey.Troshin@ihep.ru
}
\normalsize
\date{}
\maketitle

\begin{abstract}
	Transition to the reflective scattering mode which has emerged at the highest LHC energy of  $\sqrt{s}=13$ TeV  results in a relative shrinkage with the energy  of the impact parameter  region  responsible for the inelastic hadron collisions. Respective increasing role of the multiplicity fluctuations of quantum origin is emphasized. 
\end{abstract}
\end{titlepage}
\setcounter{page}{2}
\section*{Introduction}
 The  LHC measurements at the  energy of $\sqrt{s}=13$ TeV \cite{tamas} indicated that hadron scattering interaction region is evolving   to a black ring with the reflective region in the center of the ring.
 This picture corresponds to 
 the reflective scattering mode   with a peripheral impact parameter dependence of the inelastic overlap function. Such a dependence would, of course, affect observables in multiparticle production processes, and some of the related issues have already been discussed in \cite{col,eff}.  
 
 Interpretation of this mode  and its association  with a color conducting matter formation in the intermidiate state of hadron interactions has been given in \cite{jpg} and
 experimental indication on the deconfined matter formation in hadronic collisions at the LHC has been discussed in \cite{ber}.
  Theoretical  aspects  of a decoherence breaking  have been discussed in \cite{aid}.

 Here we consider  further consequences of the reflective scattering mode presence for the multiplicity distribution emphasizing the role of  the impact parameter--dependent mean multiplicity and  inelastic overlap function.  The consideration has a qualitative nature, it concerns mainly the asymptotic energy region which is at least beyond the presently available energies. But the presented conclusions are in correspondence with the observed tendencies.

 \section{Peripheral form of the inelastic overlap function and multiplicity distribution}
 Unitarity equation for the elastic scattering amplitude $F(s,t)$  has  the form
 \begin{equation} \label{unit}
 	\mbox{Im} F(s,t)=H_{el}(s,t)+H_{inel}(s,t),
 \end{equation}
where $H_{el}(s,t)$ is the two--particle intermidiate state contribution  and  $H_{inel}(s,t)$ is the sum of the  contributions from the multi--particle intermidiate states. For the forward scattering when $-t=0$   Eq. (\ref{unit}) turns into
 \begin{equation} \label{unit0}
 	\sigma_{tot}(s)=\sigma_{el}(s)+\sigma_{inel}(s),
 \end{equation}
where $\sigma_i(s)$  are the respective cross--sections. 
High--energy elastic scattering amplitude is a predominantly imaginary and is given by the sum, Eq. (\ref{unit}).
In the impact parameter representation (i.e. in the framework of quasiclassical geometrical picture, Fig. 1) the elastic and inelastic overlap functions $h_{el}(s,b)$ and $h_{inel}(s,b)$ have different profiles at high energies. 
 \begin{figure}[hbt]
	
	\hspace{1cm}	\resizebox{12cm}{!}{\includegraphics{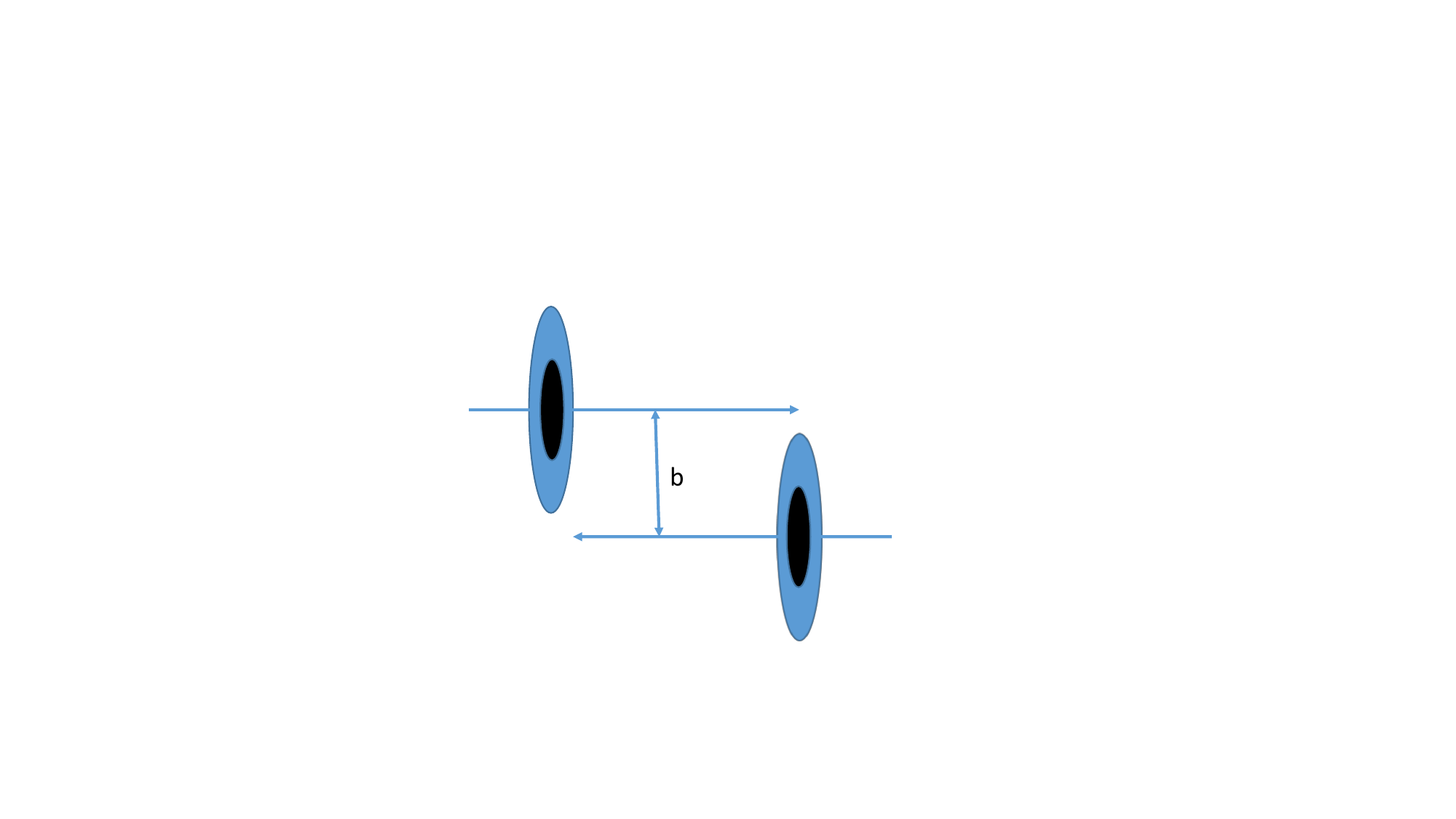}}		
	
	\caption{Schematic form of hadron scattering geometry. }	
\end{figure}	

The unitarity equation, Eq. (\ref{unit}), for the elastic scattering amplitude in the impact parameter representation, $f(s,b)$, has a diagonal form, i.e.:
\begin{equation}
\mbox{Im}f(s,b)[1-\mbox{Im}f(s,b)]=[\mbox{Re}f(s,b)]^2+h_{inel}(s,b).
\end{equation}
It is evident that  $\mbox{Re}f\to 0$ when $\mbox{Im}f\to 1$ and 
under assumption of the  vanishing real part  the following relation takes place ($f\to if$) for the inelastic overlap function $h_{inel}(s,b)$:
\begin{equation}\label{hinel}
	h_{inel}(s,b)=f(s,b)[1-f(s,b)].
\end{equation}
The impact parameter representation provides a geometric, semiclassical picture for hadron interactions. 
 The elastic overlap function preserves central profile when the energy increases. Contrary, the inelastic overlap function becomes peripheral when $f>1/2$. 
 Indeed, for $s$ and $b$ values where $f(s,b)>1/2$, the inelastic overlap function, Eq. (\ref{hinel}), decreases with the energy growth and acqures a peripheral profile (Fig. 2).  
\begin{figure}[hbt]
 
 \hspace{1cm}	\resizebox{12cm}{!}{\includegraphics{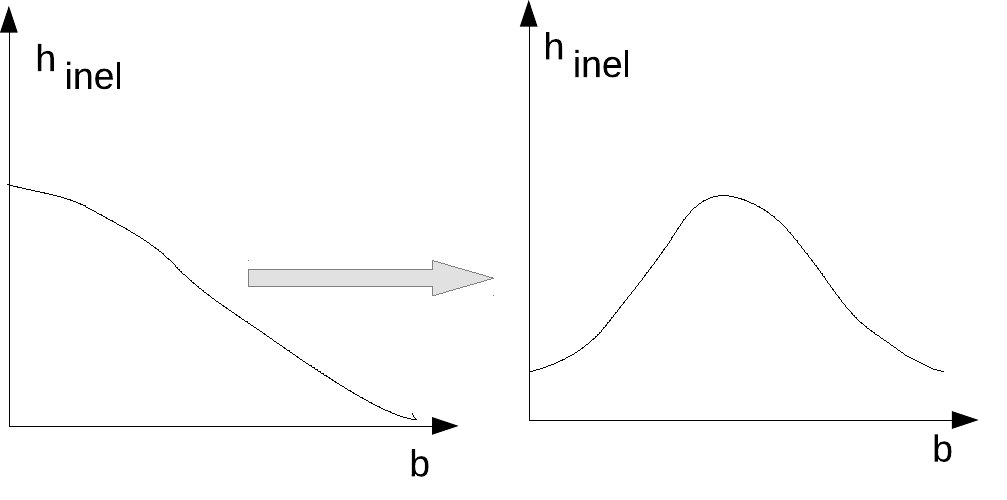}}		
 	
 	\caption{Schematic forms of inelastic overal function dependening on   impact parameter  in the shadow (left) and reflective (right) scattering modes.}	
 \end{figure}

The profile of $h_{inel}(s,b)$ becomes relatively more narrow when
$s$ increases and this concentrates our attention on the impact parameter values close to  position of the inelatic overlap function peak which we denote by $R(s)$ to keep previously used notations \cite{col,eff}. One should note that $R(s)\sim \ln s$ at      $s\to\infty$. Thus, the probability of an inelastic processes under hadron collision  at the impact parameter $b$ is
\begin{equation}
 	\sigma_{inel}(s,b)\equiv 4h_{inel}(s,b).
 \end{equation}
 with maximum at $b=R(s)$.  
 
 In what follows we use the  
 function \begin{equation}\label{pn}
 P_n(s,b)\equiv \sigma_n(s,b)/\sigma_{inel}(s,b)	
 \end{equation}
for  the multiplicity distribution at the energy $s$ and impact parameter $b$.
In Eq. (\ref{pn}), $\sigma_n(s,b)$ is the production cross--section of $n$ particles ($n\geq 3$) (see  Ref. \cite{webb} for the definitions). It seems quite logical to address the function $P_n(s,b)$ since we intend to study consequences of the reflective scattering mode and its characteristic feature – peripheral form of the inelastic overlap function $h_{inel}(s,b)$. 

The function $P_n(s,b)$ is to be used for the calculations  of the final states entropy and other thermodynamic quantities in hadron interactions. Their impact parameter dependence emphasizes an importance of the spatial proton's structure and serves as a replacement of $Q^2$--dependence of the entropy under the deep--inelastic scattering  \cite{kharz}.

 Note that 
 \begin{equation}\label{pnbb}
 P_n(s)= \int_{0}^\infty P_n(s,b)\sigma_{inel}(s,b)bdb/\int_{0}^\infty \sigma_{inel}(s,b)bdb
 \end{equation}
and 
\begin{equation}\label{nbb}
	\langle n\rangle (s)\equiv\sum _n nP_n(s)= \int_{0}^\infty \langle n \rangle(s,b)  \sigma_{inel}(s,b)bdb/\int_{0}^\infty \sigma_{inel}(s,b)bdb, 
	\end{equation}
where $\langle n\rangle (s,b)\equiv\sum _n nP_n(s,b)$ is the average final-state multiplicity at initial-state impact parameter value $b$ \cite{webb}. The latter averaging corresponds to smoothing the  quantum fluctuations of multiplicity. Those are  fluctuations of multiplicity under the fixed values of $s$ and $b$ and they originate from the probabalistic nature of the wave function of colliding protons. The recent review   and impact of the quantum fluctuations on the observales can be found in \cite{mant}.  Discussion of classical and quantum fluctuations  in Pb+Pb collisions at the LHC  was given  in \cite{olli}.

	 Quantum fluctuations are smoothed out in the function $\langle n \rangle (s,b)$ and it makes this quantity relevant for a quasiclassical modelling in the framework of impact parameter picture (section 2). In this regard  it should be noted that commutator of the impact parameter operator with the Hamiltonian is vanishing at very high energies \cite{webb} and  the impact parameter itself becomes a quasiclassical quantity.

It has been shown in \cite{eff} that the following approximate relation  between the multiplicity distribution $P_n(s)$ and  $P_n(s,b)$ occurs:
 \begin{equation}\label{flu}
 P_n(s)\simeq P_n(s,b)|_{b=R(s)}.
 \end{equation}
and  the  respective relations for  the other experimental observables such as $\langle n\rangle (s)$ are valid due to transition to the reflective scattering. The above mentioned  geometric modelling of $\langle n \rangle (s,b)$ would lead to prediction of an asymptotic energy dependence of the mean multiplicity. The   function $P_n(s,b)$ at $b=R(s)$ is : 
\begin{equation}\label{prn}
	P_n(s,b)|_{b=R(s)}= \sigma_n(s,b)|_{b=R(s)}
\end{equation}
and  the mean multiplicity $\langle n \rangle(s)$:
\begin{equation}\label{barn}
	\langle n\rangle (s)  \simeq \langle n\rangle (s,b)|_{b=R(s)}
\end{equation}
The both relations Eqs. (\ref{flu}) and (\ref{barn}) correspond to a peripheral nature of the inelastic overlap function  and imply that the averaging  goes over {\it quantum}  fluctuations of the multiplicity at   fixed impact parameter $b=R(s)$.

The distribution  $P_n (s,b)$  receives  contributions from the two sources of  different origins. These are the characteristic $b$-dependence of  
$P_n (s,b)$  associated with the varying $b$--values  and  the quantum fluctuations over $n$ at fixed values of $b$.  The function  $P_n (s,b)$ averaged over $b$, Eq. (\ref{pnbb}), results in multiplicity distribution  $P_n (s)$. Transition to mean multiplicity 
$\langle n\rangle (s)$, Eq. (\ref{nbb}), corresponds to further  averaging of   over  quantum fluctuations.

Being asymptotic, Eq. (\ref{flu})  nevertheless implies  that the event-by-event fluctuations of $n$ related to variations of the impact parameter values experiences relative weakening with the energy growth due to transition to the reflective scattering mode. Respectively, the quantum fluctuations of multiplicity become more  significant gaining an extra  weight with the energy increase.

The assumption of  an infinitely narrow distribution \cite{bron, jenk}, i.e. the delta--function dependence of muliplicity  on the impact parameter  neglects a presence of   the  quantum multiplicity fluctuations at fixed values of $b$.  Such fluctuations have been discussed  in \cite{pepin} for  the particular case of $b=0$.

 We consider now  the energy dependence of the function $\sigma_n(s,b)$ at fixed impact parameter value.  For certainty, we choose zero value of impact parameter,  $b=0$.
  The amplitude value $f(s,0)$ moves from the region of $(0,1/2]$ into the region of $[1/2,1)$, increases  and tends to unity at $s\to\infty$. Such  energy behavior of the elastic scattering amplitude corresponds to the   Chew and Frautchi pastulate  on the maximal strength of strong interactions \cite{cf,c1}. It is natural to expect that this increase is monotonic.  Respectively, the inelastic overlap function $h_{inel}(s,0)$   monotonically decreases and tends to $0$ at $s\to\infty$. It implies that $\sigma_n(s,b)$ at $b=0$ should also decrease with energy for any $n\geq 3$ since $\sigma_n (s,b)\leq \sigma_{inel} (s,b)\equiv 4h_{inel}(s,b)$, i.e. $\sigma_n (s,0)\to 0$ at $s\to\infty$ under the reflective scattering mode. In fact, this  behavior at large increasing values of energy remains to be valid for any {\it fixed} value of impact parameter: 
 \begin{equation}\label{lim}
 \lim_{s\to\infty} \sigma_n(s,b)=0.
 \end{equation}

   Quantitative  description of  multiplicity distribution in the $U$--matrix unitarization scheme (reproduces both the shadow and reflective scattering modes and naturally allows to  saturate  unitarity) was given  in \cite{rami}.

 As a result of unitarity saturation, the following relations are valid
 \begin{equation}
 \langle \Delta b^2\rangle_{inel}/\langle b^2\rangle_{inel}\sim
 \sigma_{inel}(s)/\sigma_{tot}(s)
 \sim R^{-1}(s)
 \end{equation}
   It should be noted that 
 \begin{equation}
\sigma_{tot}(s)\sim\sigma_{el}(s)\sim R^2(s)\, \mbox{and}\,   \langle b^2\rangle_{el} \sim \langle b^2\rangle_{inel}\sim R^2(s)\
\end{equation}
while
\begin{equation}\label{sinel}
  \sigma_{inel}(s) \simeq 8\pi R(s)\int_0^\infty db h_{inel}(s,b)
 \end{equation}
and the dimensional integral in Eq. (\ref{sinel})
does not depend on $s$ in the limit of $s\to\infty$ \cite{edge}, i.e. the width of the black ring does not depend on energy while the radius of the ring increases like $\ln s$. 
\section{Modelling  the mean multiplicity $\langle n\rangle (s,b)$}
As it was noted in section 1, averaging $\langle n\rangle (s,b)$  smooths the  quantum fluctuations of multiplicity and it is natural therefore to assume  correlation of the inelasticity at given impact parameter $b$ and the mean multiplicity:
\begin{equation}\label{nu}
\langle n\rangle (s,b)=\nu(s) \sigma_{inel}(s,b).
\end{equation}
It is evident  that the integrated mean multiplicity $\langle n\rangle (s)$ can be represented as the ratio:
 \begin{equation}\label{nm}
 	\langle n\rangle (s)={\nu(s)\int_0^\infty \sigma^2_{inel}(s,b)bdb}/{\int_0^\infty \sigma_{inel}(s,b)bdb}.
 \end{equation}
 The ratio of the  integrals in Eq. (\ref{nm}) is limited by unity.
 This bound takes place since $0\leq \sigma_{inel}(s,b)\leq 1$,  and asymptotically $\nu(s)\to 	\langle n\rangle (s)$. 
 
 The peripheral $b$--dependence of $\sigma_{inel}(s,b)$ under the relective scattering mode is translated into  peripheral form of $\langle n\rangle (s,b)$ and this form can be interpreted as a common consequence of the inelastic channels self--damping \cite{bak} and geometrical picture of the interaction. Effective realization of the both dynamical features is provided by the $U$--matrix unitarization scheme with relevant choice of an input \cite{rfa}. 

The importance of  the $b$--dependent  mean multiplicity $\langle n\rangle (s,b)$ studies is related to possibility of its extraction from the inclusive overlap functions introduced in \cite{sakai}. The inclusive overlap function is an overlap integral with fixed value of the  particle momentum and it becomes  the usual single--particle inclusive cross--section at $t=0$. 
It should be also noted that under  the symmetric nucleus-nucleus collisions geometrical scaling (GS)
 is observed in  $\langle n\rangle (s,b)$ dependence on centrality \cite{gs}. It would be interesting to perform similar measurements for the proton--proton collisions at the LHC energies and to test presence of  this particular GS-phenomena and validity of  Eq. (\ref{nu}), which can establish a relation  of $\langle n\rangle (s,b)$ with the  elastic scattering amplitude  due to Eq. (\ref{hinel}):
 \begin{equation}\label{nf}
	\langle n\rangle (s,b)=4\nu(s)f(s,b)[1-f(s,b)].
\end{equation} 
Eq. (\ref{nf}) implies invariance of $\langle n\rangle (s,b)$ under replacement $f\to 1-f$, i.e. it implies the same average multiplicity value corresponding to   both values:  $f$ and $1-f$. 

According to the experimental data analysis \cite{tamas}, the maximum of this function  is shifted with energy growth from $b=0$ to $b=0.4$ fm at  $\sqrt{s}=13$ TeV under assumption of Eq. (\ref{nu}).
\section*{Discussion and conclusion}
There are two scattering modes at high energies and both  are allowed  by unitarity: the shadow scattering mode (SSM) and the antishadow 
one\footnote{The term antishadowing in this context means that an increase of elastic scattering amplitude occurs under reduction of the total contribution of the inelastic channels at the given value of the impact parameter of collision} which can be interpreted as a reflective scattering mode by analogy with optics (RSM). An existence of the RSM is allowed if we do not introduce constraint $|f|\leq 1/2$.  RSM existence becomes ultimate in view of the maximal strength principle by Chew and Frautchi \cite{cf}. It is also implied by invariance of the inelastic overlap function  under replacement $f\to 1-f$
\cite{invar}. The experimental indications on the existence of this mode have been obtained in \cite{tamas}.

Gradual transition to the RSM corresponds to  the relative shrinkage  of the impact parameter variation region effectively populated by the inelastic processes. In fact, the both Eqs. (\ref{flu}) and (\ref{barn}) represent a reflection of such a behavior. 

Needless to say that  additional assumptions are needed to constrain  particular dependence of $P_n(s,b)$ on the number $n$ of produced particles ($n\geq 3$) and one cannot extract it,  in principle,  modelling  the elastic scattering amplitude only since the inelastic overlap function accounts and represents a collective effect of all the inelastic processes.

We emphasized presence of  the two different sources of multiplicity fluctuations
in hadron production at modern energies:  one is due to  variation of the collision impact parameter value  and another  one  associated with quantum fluctuations of multiplicity at fixed impact parameters.   Transition to the reflective scattering mode with the energy increase makes  the quantum fluctuations  a dominant mechanism associated with the   multiplicity fluctuations.

The models  of quantum optics    \cite{klau} can be useful for the modelling  the particle distributions at  fixed impact parameters.
The use of the gamma--distribution for the $P_n(s,b)$  in nuclei--nuclei and hadron--nuclei reactions
 has been proposed  in \cite{rog} where
discussion of the argumentation of  the impact parameter dependent multiplicity distribution   can also be found. 

Indeed, the gamma--distribution  is relevant for the  description of various systems.
Its extension for description of the small systems such as hadrons and their interactions is supported by the experimentally observed similarity of the  observables in nuclear and hadron reactions \cite{ber}, discovered the  ridge and other collective effects under interactions of small systems \cite{col, wei, cms}.     Application to  hadron collisions is complimentary gaining advantage from  validity of  the unitarity condition in this case. 
It is interesting to note that gamma--distribution  has also been applied \cite{bars} for modelling the eikonal treated as a stochastic quantity.  It transformed an original exponential form  into a rational representation of the scattering amplitude where an averaged  eikonal function serves as an input.

\section*{Aknowledgement}
We are grateful to Rami Oueslati for the interesting discussions.


\small
\begin{thebibliography}{99}
\bibitem{tamas}
T.  Cs\"{o}rg\H{o}, R. Pasechnik and A. Ster,  Acta Phys. Pol. B Proc. Suppl. {\bf 12 }, 779  (2019).
\bibitem{col}
S.M. Troshin and N.E.  Tyurin  { Int. J. Mod. Phys.  A} {\bf 26 },  4703 (2011).
\bibitem{eff}
S.M. Troshin and N.E.  Tyurin  { Int. J. Mod. Phys.  A} {\bf 29},145051 (2014).
\bibitem{jpg}
S.M. Troshin and N.E. Tyurin,  J. Phys. G {\bf 46}, 105009 (2019).
\bibitem{ber}
A. Beraudo, arXiv:2410.23750v1, contribution to LHCP 2024.
\bibitem{aid}
C.A. Aidala and T.C. Rogers, Phil. Trans. Math. Phys. Eng. Sci. {\bf 380}, 2216 (2021).
\bibitem{webb}
B.R. Webber, Nucl. Phys. B {\bf 87}, 269 (1975).
\bibitem{kharz}
Z. Tu, D.E. Kharzeev and T. Ullrich, Phys. Rev .Lett. {\bf 124},  062001 (2020).
\bibitem{mant}
H. M\"antysaari, Rept. Prog. Phys. {\bf 83}, 082201 (2020).
\bibitem{olli}
E. Roubertie, M. Verdan, A. Kirchner and  J.-Y. Ollitrault, arXiv: 2503.17035v1.
\bibitem{bron}
W. Broniowski and W. Florkowski, Phys. Rev. C {\bf 65}, 024905 (2002).
\bibitem{jenk}
L.L. Jenkovszky and B.V. Struminsky, Phys. of Atom. Nucl. {\bf 67},  48 (2004).
\bibitem{pepin}
M. Pepin, P. Christiansen, S. Munier and J.-Y. Ollitrault,  Phys. Rev. C {\bf 107},  024902 (2023).
\bibitem{cf}
G.F. Chew and S.C. Frautchi, Phys. Rev.  {\bf 123}, 1478  (1961) .
\bibitem{c1}
G.F. Chew, Rev. Mod. Phys. {\bf 34}, 394 (1962).
\bibitem{rami}
R. Oueslati and  A. Trabelsi. JHEP {\bf 07}, 100 (2024).
\bibitem{edge}
S.M. Troshin and N.E.  Tyurin,  {  Mod. Phys. Lett. A} {\bf 31},  16500025 (2016).
\bibitem{invar}
S.M. Troshin, N.E. Tyurin,  Particles {\bf 6} , 239 (2023).
\bibitem{klau}
J.R. Klauder and E.C.G. Sudarshan, {\it  Fundamentals of Quantum Optics}, W.A. Benjamin, Inc. New-York  Amsterdam 1968. 
\bibitem{rog}
R. Rogly, G. Giacalone and J.-Y. Ollitrault, Phys. Rev. C {\bf 98},  024902 (2018).

\bibitem{wei}
Wei Li, Mod. Phys. Lett. A {\bf 27}, 1230018 (2012) .
\bibitem{cms}
V. Khachatryan et al. (CMS Collaboration), Phys. Rev. Lett. {\bf 116}, 193  (2017).
\bibitem{bars}
S. Barshay, P. Heiliger and D. Rein, Mod. Phys. Lett. A {\bf 7}, 2559 (1992).
\bibitem{bak}
M. Baker and R. Blankenbecler, Phys. Rev. {\bf 128}, 415 (1962).
\bibitem{rfa}
S.M. Troshin, N.E. Tyurin,  Symmetry {\bf 14} , 1292 (2022).
\bibitem{sakai}
N. Sakai, Nuov. Cim. A {\bf 21},  368 (1974).
\bibitem{gs}
R. Rogly, G. Giacalone and J.-Y. Ollitrault, Nucl. Phys. A {\bf 982}, 355 (2019).


\end{thebibliography}
\end{document}